\documentclass{article}

\usepackage{graphicx}

\begin{document}

\newtheorem{defn}{Definition}
\def\thedefn{\thesection.\arabic{defn}}

\newtheorem{teor}[defn]{Theorem}
\newtheorem{ejem}[defn]{Example}
\newtheorem{lema}[defn]{Lemma}
\newtheorem{rema}[defn]{Remark}
\newtheorem{coro}[defn]{Corollary}
\newtheorem{prop}[defn]{Proposition}
\newtheorem{example}[defn]{Example}

\makeatother
\font\ddpp=msbm10  %at 11 truept
\def\R{\hbox{\ddpp R}}
\def\C{\hbox{\ddpp C}}
\def\L{\hbox{\ddpp L}}
\def\S{\hbox{\ddpp S}}
\def\Z{\hbox{\ddpp Z}}
\def\Q{\hbox{\ddpp Q}}
\def\N{\hbox{\ddpp N}}

\def\P{{\it Proof.} \hspace{3mm}}
\def\qed{\hfill \raisebox{1mm}{\framebox{\rule{0mm}{1mm}}}}

\newcommand{\dps}{d_{\psi}}
\newcommand{\df}{d_{\phi}}

\newcommand{\M}{{\cal M}}
\newcommand{\Mo}{{\cal M}_0}
\newcommand{\be}{\begin{equation}}
\newcommand{\ee}{\end{equation}}
\newcommand{\la}{\Lambda}
\newcommand{\dem}{\noindent {\rm Proof: }}
\newcommand{\inte}{\int_{0}^{1}}
\newcommand{\gam}{\gamma}
\newcommand{\eps}{\epsilon}
\newcommand{\<}{\langle}
\renewcommand{\>}{\rangle}
\newcommand{\Om}{\Omega^1}
\renewcommand{\(}{\left(}
\renewcommand{\)}{\right)}
\renewcommand{\[}{\left[}
\renewcommand{\]}{\right]}
\newcommand{\om}{\omega}
\newcommand{\me}{\frac{1}{2}}
\newcommand{\Mt}{\widetilde{\M}}
\newcommand{\cat}{{\mathop{\rm cat}\nolimits}}
\newcommand{\Tau}{{\cal T}}

\newcommand{\cvd}{{\rule{0.5em}{0.5em}}\smallskip}

\newcommand{\bd}{\begin{defi}}                %inizia definizione
\newcommand{\ed}{\end{defi}}                  %fine definizione
\newcommand{\bc}{\begin{coro}}                 %inizia corollario
\newcommand{\ec}{\end{coro}}                   %fine corollario
\newcommand{\bl}{\begin{lema}}                     %inizia lemma
\newcommand{\el}{\end{lema}}                       %fine lemma
\newcommand{\bp}{\begin{prop}}            %inizia proposizione
\newcommand{\ep}{\end{prop}}                %fine proposizione

\newcommand{\bere}{\begin{remark}}                  %inizia osservazione
\newcommand{\ere}{\end{remark}}                     %fine oservazione
\newcommand{\bex}{\begin{example}}            %inizia proposizione
\newcommand{\eex}{\end{example}}                %fine proposizione

\newcommand{\bt}{\begin{teor}}
\newcommand{\et}{\end{teor}}

\newcommand{\br}{\begin{rema}}
\newcommand{\er}{\end{rema}}

\newcommand{\bnr}{\begin{eqnarray*}}
\newcommand{\enr}{\end{eqnarray*}}

\newcommand{\bit}{\begin{itemize}}
\newcommand{\eit}{\end{itemize}}

\newcommand{\noi}{\noindent}

\hyphenation{Lo-rent-zian}

\title{{\bf Further results on the smoothability of Cauchy hypersurfaces and Cauchy time functions}}
\author{Antonio N. Bernal  and Miguel S\'anchez%\thanks{
%The second-named author has been partially supported by a  MCyT-FEDER Grant, MTM2004-04934-C04-%01.}
}
%\date{}
\maketitle

\begin{center}
{\small Dpto. de Geometr\'{\i}a y Topolog\'{\i}a, Facultad de Ciencias, Fuentenueva s/n, E--18071 Granada, Spain}
\end{center}

\begin{center}
Abstract
\end{center}

\noindent Recently,  folk questions on the smoothability of Cauchy
hypersurfaces and  time functions of a globally hyperbolic
spacetime $M$, have been solved. Here we give further results,
applicable to several problems:

(1) Any compact spacelike acausal submanifold $H$ with boundary
can be extended to a spacelike Cauchy hypersurface $S$. If $H$
were only achronal, counterexamples to the smooth extension exist,
but a continuous extension (in fact, valid for any compact
achronal subset $K$) is still possible.

(2) Given any spacelike Cauchy hypersurface $S$, a Cauchy temporal
function ${\cal T}$ (i.e., a smooth function with past-directed
timelike gradient everywhere, and Cauchy hypersurfaces as levels)
with $S= {\cal T}^{-1}(0)$ is constructed -thus, the spacetime
splits orthogonally as $\R\times S$ in a canonical way.

Even more, accurate versions of this last result are obtained if
the Cauchy hypersurface $S$ were non-spacelike (including
non-smooth, or achronal but non-acausal). Concretely, we construct
a {\em smooth} function $\tau: \M\rightarrow \R$ such that the
levels $S_t=\tau^{-1}(t), t\in \R$ satisfy:  (i) $S=S_0$, (ii)
each $S_t$ is a (smooth) spacelike Cauchy hypersurface for any
other $t\in \R\backslash\{0\}$. If $S$ is also acausal  then
function $\tau$ becomes a time function, i.e., it is strictly
increasing on any future-directed causal curve.

\smallskip
\smallskip

\noindent PACS 2003: 04.20.Gz ; 02.40.Ma; 04.62.+v

\smallskip


\smallskip


\section{Introduction}
The question whether the Cauchy hypersurfaces and Cauchy time
functions of a globally hyperbolic spacetime $(M,g)$ can be taken
smooth, have remained as an open folk question in Lorentzian
Geometry until the recent full solution \cite{BS1, BS2}.
Essentially, $(M,g)$ not only contains a (smooth) spacelike Cauchy
hypersurface \cite{BS1}, but also a Cauchy temporal function
${\cal T}$ and, thus, a global smooth splitting $M=\R \times S_0$,
such that each level at constant ${\cal T}$ is a Cauchy
hypersurface, orthogonal to the vector field $\partial/\partial
{\cal T}$ naturally induced from the $\R$ part \cite{BS2}
 (see
also \cite[Section 2]{BSmur} for a brief history of the problem,
and \cite{Sa} for related questions, including the smoothability
of time functions in stably causal spacetimes).  Even though such
results are enough from the conceptual viewpoint as well as for
typical applications in General Relativity (\cite{Uh}, \cite{Mu},
\cite{Bu}, etc.), the importance of spacelike Cauchy hypersurfaces
for both, classical GR (Cauchy problem for Einstein's equation)
and its quantization %(for example, \cite{Fu}), suggests some
suggests related problems.

 In the present  note,  two related problems on
 smoothability are solved. The first one concerns the
problem on extendability of a compact spacelike submanifold  to a
full smooth Cauchy hypersurface $S$. Concretely,  we will prove:

\bt \label{t} Let $(M,g)$ be a globally hyperbolic spacetime, and
let $H \subset M$ be a spacelike and  acausal compact
$m$--submanifold with boundary.

Then, there exists a spacelike Cauchy hypersurface $S$ such that
$H \subset S$. \et We emphasize that all the hypotheses are
necessary and, in fact, if $H$ were only achronal instead of
acausal, then no smooth Cauchy hypersurface extending $H$ will
exist in general (Example \ref{ex}) --even though a non-smooth one
does exist, Theorem \ref{truzz}. Moreover,  in the definitions
below we will assume that submanifolds are connected (an usual
simplifying convention)  but the proof of Theorem \ref{t} is
extended to the non-connected case easily, Remark \ref{r}.

For the second problem, recall first that the Cauchy temporal
function ${\cal T}$ in \cite{BS2} is constructed by means of a
quite long process, which uses the previously constructed Geroch's
Cauchy time function \cite{Ge}. So, even though the level
hypersurfaces  of ${\cal T}$ are proven to be Cauchy, they become
rather uncontrolled. Therefore, it is natural to wonder:
\begin{quote} Given a spacelike Cauchy hypersurface $S$, is there
a Cauchy temporal function ${\cal T}$ such that $S$ is one of its
levels?
\end{quote} Even more, if $S$ were only a topological Cauchy
hypersurface, we can wonder if it can be approximated by spacelike
ones $S_t$ in the strongest possible sense, that is, by finding a
Cauchy time function $\tau$ (everywhere smooth and with timelike
gradient except at most on $S$) such that $S_t=\tau^{-1}(t)$, for
all $ t\in \R\backslash\{0\}$, and $S=\tau^{-1}(0)$.

Our purpose is to answer affirmatively these questions. First, we
study the last one. A continuous function $\tau$ fulfilling all
the other required properties is shown to exist easily, Proposition
\ref{ap1}. By using techniques in \cite{BS1, BS2}, we show that
$\tau$ can be chosen smooth everywhere (with gradient vanishing
only at $S$, which maybe non-smooth), Theorem \ref{at1}. In this
theorem, if $S$ is acausal then $\tau$ will be automatically a
smooth time function. Nevertheless, it is not evident to prove
that, if $S$ is spacelike, then $\tau$ can be chosen with timelike
gradient at $S$. This is carried out by using again techniques in
\cite{BS2}, proving finally:

\bt \label{bt1} Let $(M,g)$ be globally hyperbolic, and $S$ a
spacelike Cauchy hypersurface. Then, there exists a Cauchy
temporal function ${\cal T}: M\rightarrow \R$ such that  $S={\cal
T}^{-1}(0)$. \et

These problems become natural as purely  geometric problems in
Lorentzian Geometry, applicable, for example, to the initial value
problem in General Relativity.  In fact, the typical initial
3-Riemannian manifold for Einstein equation would be (a
posteriori) one of the levels of a Cauchy temporal function. Thus,
from the viewpoint of foundations, one should expect   that any
spacelike Cauchy hypersurface can be chosen as one such level,
whatever the stress-energy tensor in Einstein's equation maybe
--as ensured by Theorem \ref{bt1}. Even more, some gluing methods
to solve Einstein equations make natural to extend a compact
spacelike hypersurface with boundary $H$ to a spacelike Cauchy
hypersurface $S$ under some restrictions \cite{Co, Mi}; Theorem
\ref{t} is then a fundamental claim, valid in general. On the
other hand, the interest of such a $S$ when $H$ is only a
2-surface (composed by the fixed points of a 1-parameter group of
isometries) is stressed in the report by Kay and Wald \cite{KW}.

Nevertheless, in spite of these ``classical'' applications, we are
in debt with  researchers in quantization who suggested us both
problems. Concretely, the first problem applies in locally
covariant quantum field theories  \cite{BFV}, and allows to
simplify Ruzzi's proof of the punctured Haag duality  \cite{Ru}.
The second one is useful to prove the uniqueness of solutions
to hyperbolic equations in the context of quantization.

This paper is organized as follows. In Section \ref{s2},
conventions and basic facts are recalled. The extendability of
subsets to Cauchy hypersurfaces is studied in Section \ref{s3}
from the topological viewpoint; then, smooth extensions are found
in Section \ref{s4}. In Section \ref{s5} any  Cauchy hypersurface
is proven to be the level of a Cauchy function $\tau$ which is not
only continuous but also smooth. The final improvement, i.e.,
$\tau$ can be chosen temporal if $S$ is spacelike, is carried out
in Section \ref{s6}.

\section{Set-up} \label{s2}

Throughout this paper we will use background results, notation and
conventions as in \cite{BS1, BS2}. In particular, $(M,g)$ will
denote   a spacetime, i.e., a connected time-oriented $C^k$
Lorentzian $n_0-$manifold, $n_0\geq 2, k=1,2,... \infty$; {\em
smooth} will mean $C^k$-differentiable. The signature will be
chosen $(-,+,\dots, +)$ and, thus, for a timelike (resp. causal,
lightlike, spacelike) tangent vector $v\neq 0$, one has $g(v,v)<0$
(resp. $\leq 0, =0, >0$); following \cite{O}, vector 0 will be
regarded as spacelike.
 The topological closure of a subset
$G$ is denoted as $\bar G$; if $G$ is a submanifold with boundary,
the boundary  is denoted $\partial G$.

A convex neighborhood  ${\cal C}_p$ of $p\in M$ is an open subset
which is a normal neighbourhood of all its points $q\in {\cal
C}_p$; in particular, each two points $q, q' \in {\cal C}_p$ can
be joined by a unique geodesic enterely contained in ${\cal C}_p$.
We will also consider that a complete Riemannian metric $g_R$ is
fixed and the $g_R$--diameter of any ${\cal C}_p$ is assumed to be
$<1$ (further conditions on convex neighbourhoods are frequent;
see for example {\em simple} neighbourhoods in \cite{Pe}).

Hypersurfaces and submanifolds will be always embedded, and they
will be regarded as connected and without boundary, except if
otherwise specified. In general, they are only topological, but
the {\em spacelike} ones will be regarded as smooth.

Recall that a Cauchy hypersurface is a subset  $S\subset M$ which
is crossed exactly once by any inextendible timelike curve  (and,
then, at least once by any inextendible causal curve); this
implies that $S$ is a topological hypersurface (see, for example,
\cite[Lemma 14.29]{O}). Cauchy hypersurfaces are always achronal,
but not always acausal; nevertheless,  spacelike Cauchy
hypersurfaces are necessarily acausal.

 A {\em time}
function is a continuous function $t$ which increases strictly on
any future-directed causal curve. A {\em temporal} function is a
smooth function with past-directed timelike gradient everywhere.
Temporal functions are always time functions, but even smooth time
functions may be non-temporal (as will be evident below). Time and
temporal functions will be called {\em Cauchy} if their levels are
Cauchy hypersurfaces. Notation as $S_{t_i}= t^{-1}(t_i)$ or,
simply, $S_i$ will be used for such levels.

In what follows, $(M,g)$ will be assumed globally hyperbolic. By
Geroch's theorem \cite{Ge},  it admits a {\em Cauchy time
function} $t:M\rightarrow \R$ and, from the results in \cite{BS2},
even a Cauchy temporal function ${\cal T}$ (these functions will be
assumed  onto without loss of generality). Given such a
${\cal T}$, the metric admits a (globally defined) pointwise
orthogonal splitting \be \label{euno} g=-\beta d{\cal T}^2 +
g_{\cal T}, \ee where $\beta>0$ is a smooth function on $M$, and
$g_{\cal T}$ is a Riemannian metric on each level $S_{\cal T}$.

 Given
a Cauchy hypersurface $S$, it is well known that $M$ can be
regarded topologically as a product $\R \times S$, where  $(s,x)
\ll (s',x)$ if $s<s'$ (if $S$ is smooth then $M$ is also smoothly
diffeomorphic to $\R \times S$). This result is obtained by moving
$S$ with the flow of any complete timelike vector field, but then
the slices at constant $s$ maybe non-Cauchy (even non-achronal).
Nevertheless, the aim of Theorem \ref{bt1} is to prove that (when
$S$ is spacelike) one can take $s={\cal T}$ for some temporal
function and, then, the prescribed $S$ is one of the levels of
${\cal T}$ in (\ref{euno}).

\section{Topological versions of Theorem \ref{t}} \label{s3}

For any subset $K$, we  write $J(K):= J^+(K)\cup J^-(K)$. By using
global hyperbolicity, it is straightforward to check that, if $K$
is compact then $J^\pm (K)$ and $J(K)$ are closed.

First, we will prove a topological version of Theorem \ref{t}. For this version, the following simple property becomes relevant.

\bl \label{uf}
Assume that $K\subset M$ is  compact and acausal. Then $K$ separates $J(K)$, i.e., any curve $\rho$ from $J^-(K)\backslash K$  to  $J^+(K)\backslash K$ enterely contained in $J(K)$ must cross $K$.
\el
\P Given such $\rho$, there is a point $p= \rho(s_0)$ which belongs to $J^+(K)\cap J^-(K)$, because $J^+(K)$ and $J^-(K)$ are closed. As $K$ is acausal, necessarily $p\in K$.
\cvd

If $K$ is assumed only achronal instead of acausal, then the conclusion does not hold, as the following  counterexample shows.

\bex \label{ex}
{\em  Consider the canonical Lorentzian cylinder
$(\R \times \S^1, g=-dt^2+ d\theta^2)$ and put
$K=\{(\theta/2, (\cos \theta , \sin \theta)): \theta \in [0,4\pi/3]\}$. Then, $K$ is spacelike and achronal, but not acausal because $(0,(0,0))<(2\pi/3,(-1/2, -\sqrt{3}/2))$. Even more, $J(K)$ is the whole cylinder, but $K$ does not separates it.
}
\eex

Nevertheless, given such an achronal  $K$, a new achronal set $K^C \supset K$ can be constructed which satisfies  the conclusion in Lemma \ref{uf}. Concretely, recall that if $K$ is achronal but not acausal then there are points $p, q\in K$, $p<q$ which will be connectable by a lightlike geodesic segment $\gamma_{pq}$ (without conjugate points except at most the extremes). Define the {\em causal hull} of $K$ as the set $K^C \subset M$ containing $K$ and all the lightlike segments which connect points of $K$. Obviously, $J(K)=J(K^C)$ and, if $K$ is compact and achronal, then so is $K^C$. A straightforward modification of Lemma \ref{uf} yields:

\bl \label{uf2}
Assume that $K\subset M$ is  compact and achronal. Then the causal hull $K^C$ separates $J(K)$.
\el

The topological version of Theorem \ref{t} is the following one (recall that $K$ does not need to be a submanifold here):

\bp \label{pruzz}
Let $(M,g)$ be a globally hyperbolic spacetime, and let $K\subset  M$ be an  acausal (resp. achronal) compact subset.

Then, there exists an acausal Cauchy (resp. a Cauchy) hypersurface $S'_K$ such that $K \subset S'_K$.
\ep

\P We will reason the acausal case; the achronal one is analogous, replacing along the proof $K$ by $K^C$ and using Lemma \ref{uf2} instead of Lemma \ref{uf}.

Notice first that, as $J(K)$ is  closed,  its complement $M'= M\backslash J(K)$, if non-empty, is an open subset of $M$ which, regarded as a (possibly non-connected) spacetime, becomes globally hyperbolic too. In fact, $M'$ is obviously strongly causal and,  given any $p, q \in M'$, the compact diamond  $J^+(p) \cap J^-(q) \subset M$ is included in $M'$.

Now, consider any (possibly non-connected) acausal Cauchy hypersurface $S'$ of $M'$ (if $M'=\emptyset$, put $S'=\emptyset$)  and let us check that the choice $S'_K= S' \cup K$ is the required hypersurface. Recall first that $S'_K$ is an acausal subset of $M$, because if a (inextendible, future-directed) causal curve $\gamma$ crosses $K$ then it is completely contained in $J(K)$ and cannot cross $S'$.

On the other hand, if $\gamma$ is not completely contained in $J(K)$, it crosses obviously $S'$. Otherwise, it will cross $K$ because if, say, $\gamma(s_0) \in J^+(K)$, then $Z=J^-(\gamma(s_0)) \cap J^+(K)$ is compact and, as $\gamma$ cannot remain imprisoned towards the past in $Z$, it will reach $J^-(K)$, and Lemma \ref{uf} can be claimed. \cvd

\begin{rema} \label{rh} {\rm
The compactness of $K$ becomes essential in Proposition \ref{pruzz}, as one can check by taking $K$ as the upper component  of hyperbolic space (or closed non-compact subsets of it) in Lorentz-Minkowski spacetime.
}\end{rema}

The following technical strenghtening of Proposition \ref{pruzz} shows (also at a topological level) that the Cauchy hypersurface can be controlled further, inside the levels of a Cauchy time function.

\bt \label{truzz}
Fix a Cauchy time function $t: M\rightarrow \R$ in $(M,g)$. If $K$ is a compact acausal (resp. achronal) subset
%of a (acausal) Cauchy hypersurface $S_K$
then there exists $t_1<t_2$ and an acausal Cauchy (resp. a Cauchy)  hypersurface $S_K \supset K$ such that:
\be \label{eruzz}
S_K \subset J^+(S_1) \cap J^-(S_2).
\ee
\et

\P Again, we will consider only the acausal case. From Proposition \ref{pruzz}, we can assume that $K$ is a subset of an acausal Cauchy hypersurface $S'_K$.

Let $t_1$ (resp. $t_2$) be the minimum (resp. maximum) of $t(K)$, choose any $t_0 \in \R$, and
regard $S_1, S_2, S'_K$ as graphs on $S_0$, i.e.:
$$
S'_K=\{(t'_K(x),x): x\in S_0\}
$$
for some continuous $t'_K: S_0 \rightarrow \R$, and analogously
for $S_i$ (each function $t_i(x)$ is constantly equal to $t_i$,
for $i=1,2$). Define $t_K:S_0\rightarrow \R$ as: \be
\label{egraph}
t_K(x) = \left\{
\begin{array}{lll}
t'_K(x) & \hbox{if} & t_1 \leq t'_K(x) \leq t_2 \\
t_1 & \hbox{if} &  t'_K(x) < t_1 \\
t_2 & \hbox{if} &  t_2 < t'_K(x)
\end{array}
\right. \ee Clearly, $t_K$ is continuous
($t_K=$Max$(t_1,$Min$(t_2,t'_K))$), and the corresponding graph
$S_K$ is a closed topological hypersurface which includes $K$. To
check that $S_K$ is Cauchy, recall first that it is crossed by any
inextendible timelike curve $\gamma$ (as $\gamma$ must cross $S_1$
and $S_2$). It is also achronal because, if $p,q \in S_K$ were
connectable by means of a future-directed causal curve $\gamma$,
then at least one of them is not included in $S'_K$, say,
$p=(t_1,x_1)\in S_1$. Recall that, as $t'_K(x_1) < t_1$, then
$S'_K \ni (t'_K(x_1),x_1) \ll p < q$, and the acausality of $S'_K$
forces $q\not\in S'_K$. Thus $q=(t_2,x_2)\in S_2$  but then
$(t'_K(x_1),x_1) \ll p < q \ll (t'_K(x_2),x_2)\in S'_K$, in
contradiction again with the acausality of $S'_K$. \cvd

Recall from Example \ref{ex} that, in general, one  cannot hope to
extend a spacelike achronal compact hypersurface to a smooth
Cauchy hypersurface. Thus, Theorem \ref{truzz} is our best result
for this case\footnote{The proofs of Theorem \ref{truzz} and
Proposition \ref{pruzz} also show that, if $K$ is assumed only
achronal, the points in $S_K$ where the acausality is violated
belongs to the convex hull $K^C$.}. Nevertheless, in the next
section, we will see that the acausal ones can be smoothly
extended. The following simple result for the acausal case, also
suggest the main obstruction for the achronal one (again, the
conclusion would not hold in Example \ref{ex}).

\bl \label{l0} In the hypotheses of Th. \ref{t}, there exist two
hypersurfaces  $G_1, G_2$ such that the closures $\bar G_1, \bar
G_2$ are acausal, compact, spacelike hypersurfaces with boundary,
and $H \subset G_1$,  $\bar G_1 \subset G_2$. \el

\P To  prove the existence of $G_1$ is enough. It is
straightforward to prove the extendability of the compact
$k$-submanifold with boundary $H$ to an hypersurface $G_1$ which
can be chosen spacelike by continuity (and with $\bar G_1$
compact). Let us prove that $\bar G_1$ can be also chosen acausal.

Otherwise, taken any sequence of such hypersurfaces with boundary $\{\bar G_n\}_n$ %(starting in $\bar G_1$)
with $\bar G_{n+1} \subset \bar G_{n}$, and $H = \cap_n \bar G_n$,
then no hypersurface $\bar G_n$  would be acausal. Thus, we can
construct two  sequences $p_n < q_n$, $p_n, q_n \in \bar G_n$ (and
the corresponding sequence of connecting causal curves $\gamma_n$)
which, up to subsequences, converge to limits $p,q$ in $H$.
%$\{p_n\}\rightarrow p$, $\{q_n\}\rightarrow q$,  .
If $p\neq q$ this would contradict the acausality of $H$ (the
sequence $\{\gamma_n\}_n$ would have a causal limit curve);
otherwise, the strong causality of $(M,g)$ would be violated at
$p$. \cvd

\section{The smoothing procedure for Theorem \ref{t}} \label{s4}

In what follows, the notation and ambient hypotheses of Theorem
\ref{t} and Lemma \ref{l0} are assumed, and $S_0=S_{t_0}$ for a
choice $t_0<t_1$. $S_K$ will denote the Cauchy hypersurface in
Theorem \ref{truzz} obtained for $K=\bar G_2$. Our aim is to
smooth $S_K$ outside $H$.

\br {\em Notice that, by using the smoothness results \cite{BS1, BS2}, the Cauchy
hypersurface $S'_K$ in Proposition \ref{pruzz} is the graph of a function
$t'_K$,  which can be assumed smooth everywhere but the points
corresponding to $\partial G_2$. This suggests the possibility to
obtain the required smooth hypersurface by smoothing $t'_K$. Nevertheless,
the limits of the derivatives of $t'_K$ as one approaches
$\partial G_2$ from outside are widely uncontrolled (say, for a
sequence $\{p_n\}_n \subset M\backslash \bar G_2$ converging to a
point $p\in \partial G_2$ the sequence of --spacelike-- tangent
spaces $T_{p_n}S'_K$ may converge to a degenerate hyperplane of
$T_pM$). Thus, we prefer to follow systematically the approach in
\cite{BS1}. }\er

We will need first the following two technical results. The first
one was proved in \cite{BS1}:

\begin{lema} \label{local}
Let $G_0$ be an open neighbourhood of  $H$ in $G_1$, fix $p \in
S_{K}\backslash G_0$, and any convex neighborhood of $p$, ${\cal
C}_p \subset I^+(S_{0})$ which does not intersect $H$.

Then there exists a smooth function
$$
h_p: M \rightarrow [0, \infty)
$$
which satisfies:

(i) $h_p(p)=2$.

(ii) The support of $h_p$  is compact and included in ${\cal C}_p$.

(iii) If $q \in J^-(S_{K}) $ and $h_p(q) \neq 0$ then $\nabla h_p(q)$ is timelike and past-directed.
\end{lema}

\noindent In fact, we can take $h_p(q)= a_p e^{-1/d(q,p')^2}$ where $p'\in I^-(p)$ is chosen close to $p$ and $a_p$ is a constant of normalization (see \cite[Lemma 4.12]{BS1} for details).   The second one can be seen as a refinement of previous
lemma, in order to deal with the submanifold $H$ instead of a
point $p$.
\begin{lema} \label{l1}
There exists a function
$$
h_G: M \rightarrow [0, \infty)
$$
which is smooth on (a neighborhood of) $J^-(S_K)$ and satisfies:

(i)  %$h_G((G_1) \equiv 1$
$h_G(q) = 1$ for all $q$ in some neighbourhood $ G_0$ of $H$ in
$G_1$; $h_G(S_K\backslash G_2) \equiv 0$.

(ii) The support of $h_G$ (i.e., the closure of $h_G^{-1}(0,
\infty)$)  is included in $I^+(S_0)$ for some $t_0\in \R$.

(iii) If $q \in J^-(S_{K})$ and $h_G(q) \neq 0$ then $\nabla h_G(q)$ is timelike and past-directed.% (and, by continuity, this also holds in some neighbourhood of $H_1$).
\end{lema}

\P Essentially,  $h_G$ will be  the time-separation function (up
to a normalization and a suitable  exponentiation, to make it
smooth at 0) to a hypersurface $\tilde G$ ($= G^\epsilon$ for
small $\epsilon$) which behaves as in the figure.

\begin{figure}[ht]
\centering
\includegraphics[width=12cm]{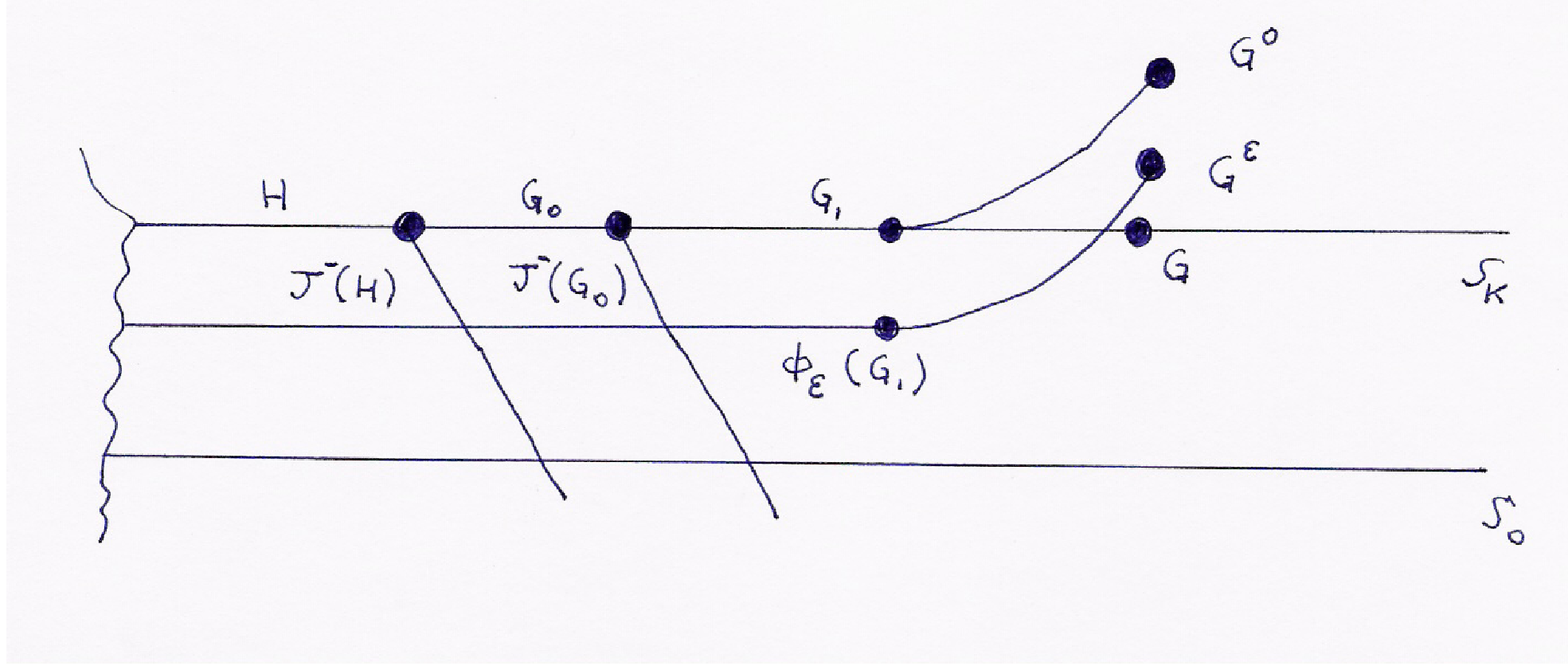}
\caption{$\bar G^0$ is a spacelike graph on $\bar G$ ($H\subset
G_1; \bar G_1 \subset G$).  Moreover, $\bar G_1\subset  G^0$ and,
thus, $\phi_\epsilon (G_1) \subset \phi_\epsilon(\bar G^0)$. For
small $\epsilon$, $J^-(G_0)\cap \bar G^\epsilon \subset
\phi_\epsilon(G_1)$ and $d(\bar G^\epsilon, p)=\epsilon$ for all
$p\in G_0$.  $\tilde G$ is taken equal to such $G^\epsilon$. }
\label{fig1}
\end{figure}

Rigourously, let $N$ be the unitary future-directed normal vector
field to $G_2$, and let $d_R$ be the distance associated to the
induced Riemannian metric on $G_2$. Denote as $d_R(\cdot, \bar
G_1)$ the $d_R$-distance function to $\bar G_1$ on $\bar G_2$, and
let
$$\rho: \bar G_2 \rightarrow [0,\infty), \quad  \rho(p) =
\exp(-d_R(p,\bar G_1)^{-2}). $$ Note that this function is $0$ on
$\bar G_1$, and smooth on some open neighbourhood $G$ of $\bar
G_1$. Even more, taking if necessary a smaller $G$, we can assume
that the graph
$$\bar G^0 =\{\exp(\rho(p)N_p) : p\in \bar G \}$$
is a spacelike compact hypersurface with boundary.

Let $\epsilon_0>0$ be the minimum of $\rho$ on the boundary
$\partial G$, and let $$\Phi: [0,\delta_0) \times \bar G
\rightarrow M, \quad \Phi(s,p)=\exp(-sN_p),$$ where
$\delta_0<\epsilon_0$ is small enough to make $\Phi$ a
diffeomorhism onto its image (see, for example, \cite[Section
7]{Pe} or \cite[Ch. 10]{O} for background details). Now, for any
$\epsilon \in (0,\delta_0)$ the hypersurface
$$\bar G^\epsilon =\{\Phi(\epsilon,p)= \exp(-\epsilon N_p) : p\in \bar G^0 \}$$
satisfies:

(a) $\partial G^\epsilon \subset I^+(S_K)$

(b) there are no focal points of $G^\epsilon$ in $I^+(G^\epsilon)
\cap J^-(S_K)$.
\smallskip

\noi Therefore, the time-separation function (Lorentzian distance)
to $\bar G^\epsilon$, $d(\cdot, \bar G^\epsilon)$, is smooth on a
neighbourhood of $I^+(G_\epsilon) \cap J^-(S_K)$, with
past-directed timelike gradient. Even more, choosing $\epsilon$
small enough we have, for some neighbourhood $G_0$ of $H$:

(c) $J^-(G_0) \cap G^\epsilon \subset \Phi(\epsilon, G_1)$ and,
thus, $d(p,G^\epsilon)=\epsilon$, for all $p \in G_0$.

\smallskip

\noi Now, the required hypersurface $\tilde G$ is any such
$G^\epsilon$ satisfying (a), (b), (c), and $h_G$ is then:
$$ h_G(p)= \exp(\epsilon^{-2}-d(p, \tilde G)^{-2}). $$
%being extended to $J^+(S_K)$ in any smooth way.
\cvd

\noindent With the two previous technical lemmas at hand, the following modification of \cite[Prop. 4.14]{BS1} can be carried out.

\bp \label{global}
There exists a smooth function
$$
h: M \rightarrow [0, \infty)
$$
which satisfies:

%\begin{itemize}
%\item[1. ]

(i) $h(q) \geq 1$ if $q \in S_K$, with $h(H)\equiv 1$,
$h(S_K\backslash G_0) > 1$.

%\item[2. ]

(ii) $h(q)=0$ if $q\in J^-(S_0)$.

%\item[3. ]

(iii) If $q \in J^-(S_{K})$ and $h(q) \neq 0$ then $\nabla h(q)$ is timelike and past-directed.

%\end{itemize}
\ep \dem Consider for any $p\in S_K\backslash G_0$ the convex
neighborhood ${\cal C}_p$ in Lemma \ref{local}, and take the
corresponding function $h_p$. Let $W_p =
h_p^{-1}\left((1,\infty)\right)$ ($W_p \subset {\cal C}_p$), and
$W_G=h_G^{-1}(0,\infty)$. Obviously,
$${\cal W} = \{ W_p, \; p\in  S_K\backslash G_0\} \cup \{W_G\} $$
covers the closed hypersurface $S_K$, and admits a locally finite
subcovering\footnote{The assumption on the bound by 1 of the
$g_R$-diameter of ${\cal C}_p$ (Section \ref{s2}) is used here;
see \cite[Lemma 4.13]{BS1} for related details.} ${\cal W}' = \{
W_{p_i}, \; i\in \N \} \cup \{ W_G \}$. Notice that $W_G$ must be
included necessarily in the subcovering, because $H \cap W_p
=\emptyset$ for all $p$. Then, it is easy to check that
\be \label{ez} h = h_G + \sum_i h_i. \ee
fulfills all the requirements.
\cvd

{\em Proof of Theorem \ref{t}.}

Notice that, by continuity, property (iii) of Proposition \ref{global} also
holds in some neighbourhood $V$ of $S_K$. Thus, $h$ admits 1 as a
regular value in $V\cup J^-(S_{K})$ and the connected component
$S$ of $h^{-1}(1)$ which contains $H$ is included in the closed
subset $J^-(S_{K})$. Thus, $S$ is a closed spacelike hypersurface
of $M$ which lies between two Cauchy hypersurfaces $S\subset
J^+(S_0)\cap J^-(S_K)$ and, then, it is itself a Cauchy
hypersurface \cite[Corollary 3.11]{BS1}. \cvd

\br \label{r} {\em If $H$ where not connected  (but still
acausal), then the proof would remain essentially equal, just
considering in (\ref{ez}) the sum of  functions $h_G^j, j=1,\dots
l$ for each  connected component $H^j$ of $H$ (instead of only one
$h_G$). Notice that $h^{-1}(1) \cap J^-(S_{K})$ cannot have more
than one connected component and, thus, a hypersurface containing
all the $H^j$'s is obtained. In fact, the connectedness of
$h^{-1}(1) \cap J^-(S_{K})$ follows because otherwise each
connected component would be a Cauchy hypersurface. But if there
were more than one component, a timelike curve $\gamma$ in
$J^-(S_K)$ would cross two such Cauchy hypersurfaces, in
contradiction with property (iii) ($h$ increases on $\gamma$ but
if it is constant on $h^{-1}(1) \cap J^-(S_{K})$).}\er

\section{Continuous and smooth versions of Theorem \ref{bt1}} \label{s5}

 Along this section $S$ will be one of the topological
Cauchy hypersurfaces of $(M,g)$ (non-necessarily smooth nor
acausal). If $U\ni p$ is a neighbourhood, $J^+(p,U)$ denotes the
causal future of $p$ computed in $U$, regarded $U$ as a spacetime.

Our aim will be to prove:

\bt \label{at1} Let $(M,g)$ be globally hyperbolic, and $S$ a
Cauchy hypersurface. There exists a smooth onto
 function $\tau: M\rightarrow \R$ such that:

(i) $S=\tau^{-1}(0)$.

(ii) The gradient  $\nabla \tau$ is past-directed timelike on
$M\backslash S$.

(iii) Each level $S_t=\tau^{-1}(t), t\in \R\backslash\{0\}$ is a
spacelike Cauchy hypersurface.

\smallskip

%\noindent
Even more, if $S$ is acausal then $\tau$ is also a smooth time
function. \et

\bl \label{al1} $I^+(S)$ and $I^-(S)$, regarded as spacetimes, are
globally hyperbolic. \el

\dem Straightforward by taking into account (for $I^+(S)$):
$$J^+(p, I^+(S)) \cap J^-(q, I^+(S)) = J^+(p) \cap J^-(q) , \quad \quad \forall p, q \in I^+(S).$$
\cvd

\noi Our construction of the function with the required properties
(along this section and the next one), starts at the following
result, which will be subsequently refined: \bp \label{ap1} There
exists a continuous onto function $\tilde \tau: M\rightarrow \R$
such that:

(i) $S=\tilde \tau^{-1}(0)$.

(ii) $\tilde \tau$ is smooth with (past-directed) timelike
gradient on $M\backslash S$.

(iii) Each  $S_t , t\in \R\backslash\{0\}$ (inverse image of the
regular value $t\neq 0$) is a spacelike Cauchy hypersurface.

\smallskip

%\noindent
Even more, if $S$ is acausal then $\tilde \tau$ is also a time
function. \ep

\dem Let ${\cal T}_{S^+}$ (resp. ${\cal T}_{S^-}$) be a (onto)
Cauchy temporal function on $I^+(S)$ (resp. $I^-(S)$), as
constructed in \cite{BS2}. Define
$$
\tilde \tau(p)= \left\{
\begin{array}{ll}
\exp({\cal T}_{S^+}(p)), & \forall p\in I^+(S) \\
0, & \forall p\in S \\
-\exp(-{\cal T}_{S^-}(p)), & \forall p\in I^-(S).
\end{array}
\right.
$$
%It is not difficult to check that $\tilde \tau$ satisfies all the
%properties.
In order to check the continuity of $\tilde \tau$ on $S$, take any
sequence $\{x_k\}\rightarrow x_0\in S$. Assume $\{x_k\}_k\subset
I^+(S)$ (the case $I^-(S)$ is analogous), and recall that all the
points $x_k$ must lie in a compact neighbourhood $W$ of $x_0$.
Moreover, each $x_k$ belongs to a level $\tilde S_{t_k}= \tilde
\tau^{-1}(t_k)$ of $\tilde \tau$ (and ${\cal T}_{S^+}$). If
$\{t_k\}_k$ did not converge to $0$ then, up to a subsequence,
$\{t_k\}\rightarrow t_0>0$ and the limit $x_0$ of $\{x_k\}$ would
lie in $W\cap S_{t_0} \subset I^+(S)$, a contradiction.

To check that  $\tilde \tau$ satisfies the remainder of required
properties  becomes straightforward.

 \cvd

In the  proof above, the levels of $\tau$ are clearly Cauchy
hypersurfaces. Nevertheless, the analogous step in the proof of
Theorem \ref{at1} will not be so evident, and the following result
will be useful (see \cite[Corollary 11]{BS1}  or \cite[Corollary
2]{Ga}):
\begin{prop} \label{ap2} Let  $S_{-}$ and $S_+$ be two disjoint Cauchy hypersurfaces of $M$ with $S_{-} \subset I^-(S_+)$. Then any  connected,
closed (as a topological subspace of $M$), spacelike hypersurface
contained in $I^+(S_{-}) \cap I^-(S_+)$ is a Cauchy hypersurface.
\end{prop}

\smallskip

\noi {\em Proof of Theorem \ref{at1}.}

Notice that the function $\tilde \tau$ constructed in Proposition
\ref{ap1} satisfies all the conditions but the smoothability at
$S$. In order to obtain this, consider a sequence of smooth
functions $\varphi^+_k, \varphi^-_k: \R  \rightarrow  \R$ such
that:
$$
\varphi^+_k(t) = \left\{
\begin{array}{ll}
0, & \forall t \leq 1/k
 \\
t, & \forall t \geq 2
\end{array}
\right. \quad \quad \varphi^-_k(t) = \left\{
\begin{array}{ll}
t, & \forall t \leq -2 \\
0, & \forall t \geq -1/k
\end{array}
\right.
$$

$$
\frac{d\varphi^+_k}{dt}(t_0)>0, \; \forall t_0> 1/k, \quad \quad
\frac{d\varphi^-_k}{dt}(t_0)>0, \; \forall t_0< -1/k.
$$
Choose constants $C_k\geq 1$, %$m=1, 2, \dots$
such that: \be \label{ae1}
\left|\frac{d^m\varphi^+_k}{dt^m}(t_0)\right|< C_k, \quad
\left|\frac{d^m\varphi^-_k}{dt^m}(t_0)\right| < C_k, \quad \forall
m\in \{1,\dots,k\}. \ee Now, put:
$$
\tau_k^+ = \varphi^+_k \circ \tilde \tau , \quad \quad \tau_k^- =
\varphi^-_k \circ \tilde \tau .
$$
Recall that $\nabla \tau_k^+$, (resp. $\nabla \tau_k^-$) is
timelike past-directed in $I^+(S_{1/k})$ (resp. $I^-(S_{-1/k})$)
and 0 otherwise. The required function is then:
$$
\tau= \Lambda \sum_{k=1}^\infty \frac{1}{2^k C_k} \left(\tau_k^+ +
\tau_k^-\right)
$$
with $\Lambda=\sum_k 2^k C_k$. In fact, the smoothness of $\tau$
follows from the definition of $\tau_k^\pm$ and the bounds in
(\ref{ae1}) (see \cite[Theorem 3.11]{BS2} for related
computations). Property (i) is trivial, and (ii) follows from the
convexity of the timecones (recall \cite[Lemma 3.10]{BS2}). For
(iii), recall that $\tau^{-1}(t)= \tilde\tau^{-1}(t)$ for $|t|\geq
2$; for the case $|t|<2$, use Proposition \ref{ap2} with
$S_-=S_{-2}, S_+=S_{2}$. \cvd

\section{From smooth time functions to temporal functions}
\label{s6}

 Along this section the Cauchy hypersurface $S$ will be
spacelike, and we will fix a smooth time function $\tau$ as in
Theorem \ref{at1}. Our purpose is to complete the proof of Theorem
\ref{bt1}.

\bl \label{bl1} Let $W \subset \tau^{-1}(-1, 1)$ be an open
neighborhood of $S$.  Then, there exist a function $h_+$ (resp
$h_-$) on $M$ such that:

(i)  $h_+\geq 0 $ (resp. $h_-\leq 0$) on $M$, and $h_+ \equiv 0$
on $I^-(S)\backslash W$ (resp. $h_- \equiv 0$ on $I^+(S)\backslash
W$).

(ii) $h_+ \equiv \, 1$ on $J^+(S)$ (resp. $h_- \equiv \, -1$ on
$J^-(S)$).

(iii) If $\nabla h_+(p)$ (resp. $\nabla h_-(p)$) does not vanish
at $p \, (\in W )$ then $\nabla h_+(p)$ (resp. $\nabla h_-(p)$) is
timelike past-directed. \el

\dem For $h_-$, take the function $h^-$ in \cite[Lemma 3.8]{BS2}
with $U=W \cup J^-(S)$ (function $h_+$ can be constructed
analogously). \cvd

\noi Taking into account the notion of ``time step function'' in
\cite{BS2}, the key step now is to prove the existence of a
function $\tau_0$ as in \cite[Proposition 3.6]{BS2} such that $S$
is a level of $\tau_0$. More precisely: \bp \label{bp2} Given the
Cauchy hypersurfaces $S_i = \tau^{-1}(i), i=-1,0,1$ ($S\equiv
S_0$) there exists a function $\tau_0$ such that:

\begin{enumerate}

\item $\nabla \tau_0$ is timelike and past-directed where it does
not vanish, that is, in the interior of its support $V :={\rm
Int(Supp}(\nabla \tau_0 )$).

\item $-1 \leq \tau_0 \leq 1$.

\item $\tau_0(J^+(S_{1})) \equiv 1$, $\tau_0(J^-(S_{-1})) \equiv
-1$. In particular, $V \subset \tau^{-1}(-1,1).$

\item $S = \tau_0^{-1}(0) \subset V$.

\end{enumerate}
\ep \dem Consider the function $p\rightarrow d(p, S)$, where
$d(\cdot,S)$ is the signed distance to $S$, i.e,  $|d(p,S)|$ is
the maximum length of the causal curves from $p$ to $S$, $d(p,S)
\geq 0$ if  $p\in J^+(S)$, and $d(p,S) < 0$ otherwise.  The
function $h_S(p)= d(p,S)+1$ satisfies  $h_S(S)\equiv 1$ obviously.
Moreover, $h_S$ is smooth, positive and with past-directed
timelike gradient in the closure of some open neighbourhood $W$ of
$S$. Without loss of generality, we will assume $W\subset
\tau^{-1}(-1,1)$.

Now, consider the function $h_+$ provided in Lemma \ref{bl1}.
Clearly,
$$ h^+= h_S h_+$$
is smooth, non-negative and well defined in $W \cup J^-(S)$
(putting $h^+ \equiv 0$ on $J^-(S)\backslash W$). Even more,
$\nabla h^+$ is timelike and past-directed where it does not
vanish (simply, use Leibniz's rule and the conditions imposed on
$h_+, h_S$; notice that the role of $h^+$ is similar to the
function also labelled $h^+$ in \cite[Proposition 3.6]{BS2}).
Taking $h_-$ from Lemma \ref{al1}, the required function will be
$$
\tau_0 = 2 \; \frac{h^+}{h^+ - h_-} -1
$$
($\tau_0$ is extended as 1 on $J^+(S)\backslash W$). In fact,
notice that $$ \nabla \tau_0 = 2 \; \frac{h^+ \nabla h_- - h_-
\nabla h^+}{(h^+ - h_-)^2}
$$
is either timelike or 0 everywhere, and also the other required
properties hold. \cvd

\noindent {\em Proof of Theorem \ref{bt1}.}

The required function is
$$ {\cal T}= \tau + \tau_0,$$
where $\tau$ is taken from Theorem \ref{at1} and $\tau_0$ from
previous proposition. In fact, ${\cal T}$ is obviously a temporal
function with $S={\cal T}^{-1}(0)$. Even more, the levels of
${\cal T}$ are Cauchy hypersurfaces: ${\cal
T}^{-1}(t)=\tau^{-1}(t)$ for $|t|>1$ and, for the case $|t|<1$,
use Proposition \ref{ap2} with $S_-=S_{-1}, S_+=S_{1}$. \cvd

\section*{Acknowledgments}

The problem studied in Theorem \ref{t} and its motivations were
suggested to us by Professors Romeo Brunetti (Univ. Hamburg) and
Giuseppe Ruzzi (Univ. of Rome ``Tor Vergata"). The problem in
Theorem \ref{bt1} was posed by Professors Christian B\"ar,
Nicholas Ginoux and Frank Pf\"affle  (Univ. Potsdam). We
acknowledge warmly their encouragement and support to all of them.
These problems were also discussed in the interdisciplinary
meeting ``Hyperbolic operators and quantization'' (Erwin
Schr\"odinger Institute, Vienna,  November 2005); the support of
organizers and participants is also acknowledged.

The second-named author has also been partially supported by a
Spanish MCyT-FEDER Grant, MTM2004-04934-C04-01, as well as an
Andalusian regional Grant, FQM 324.

\end{document}